# Earthquake Predicting and Tectonic Stress Reducing Possibilities

Manana Kachakhidze[*,1], Nino Kachakhidze-Murphy[1]

1Georgian Technical University, Natural Hazard Scientific-Research Center
0171, Georgia, Tbilisi, Kostava str. 77
[*]Corresponding author: Manana Kachakhidze kachakhidzem@gmail.com

## Abstract

*Recent satellite and ground-based observations prove that during the earthquake formative period VLF/LF and ULF electromagnetic emissions are observed in seismogenic areas.*

*The present consolidated paper describes a theoretical model of the generation of electromagnetic emissions detected during the earthquake preparation periods, a scheme of the earthquake prediction methodology, the possible methods of the large earthquake prediction on the basis of the European Network data.*

*The offered methods are capable of simultaneous determination of all three characteristic parameters necessary for incoming earthquake prediction.*

*Today existing networks, detect this electromagnetic radiation, consisting of stationary transmitters and receivers. Since in some cases these networks are unable to detect any electromagnetic radiation, we have suggested the possible arrangements of the VLF/LF mobile network.*

*The paper offers a completely different view of seismic risk assessing and reducing the accumulated tectonic stress in the earthquake focus as well.*

## Introduction

Earthquake hazards studies have intensified worldwide during the second half of the last century. Besides theoretical studies, it has become possible to conduct field, laboratory, and satellite experiments. These studies have revealed various geophysical fields anomalous changes in the lithosphere-atmosphere-ionosphere system, prior to the earthquakes. These anomalies may accompany the earthquake preparation process and emerge several months, weeks, or days before an earthquake.

Among geophysical anomalies, specific attention has been given to the electromagnetic emissions (very low frequency (VLF), low frequency (LF), and ultralow frequency (ULF)) recorded before earthquakes, with an interesting correlation between seismic activity and disturbances to radiobroadcasts (1, 2, 3, 4, 5, 6, 7, 8, 9, 11,15, 20, 21, 23, 32, 35, 42, 44, 45).

To collect VLF/LF radio signals and study the electromagnetic field variations associated with seismogenic processes, since the 2000s VLF radio networks were established by Japanese researchers (Japanese-Pacific VLF/LF Network) (22) and a European network (5,6) by Japanese, Russian, and Italian teams cooperation. It is worth underling that there are examples of large earthquakes when no radiation has been detected (4,8).

This has led the scientific community to doubt that electromagnetic emissions may not always occur during earthquake preparation.

But this contradicts the long-established fact that electromagnetic radiation is generated during crack formation, which has been illustrated both theoretically and experimentally (9,10,11,13,14,15,16,18, 41, 42, 43, 44). We agree with the position of scientists investigating electromagnetic emissions in the field and laboratories for many years and who concluded that VLF/VHF electromagnetic precursors do exist.

The development of suitable observational techniques and analysis methods is, therefore, a promising research direction for the study of earthquake precursors (12).

## Results:

1. **Theoretical model**

    Based on electrodynamics it is created a theoretical model of the generation of electromagnetic emissions detected prior to earthquakes.

2. **Earthquake prediction methodology and methods**

    A possible methodology for earthquake prediction is created by a model of electromagnetic radiation generation and a geological model of fault formation.

    Besides, by analyses of electromagnetic radiation retrospective data, the large earthquake prediction methods are formulated.

    VLF/LF electromagnetic radiation frequency changing describes the full process of main fault formation and therefore it turned out to be the unique precursor, capable of large, inland earthquake prediction.

3. **Mobile network and risk reduction**

    The advantage of an earthquake precursor mobile network arrangement is shown.

    A new vision of reducing the risk of earthquakes by artificially decreasing tectonic stress is formulated.

## Discussion

### Theoretical Model of electromagnetic emissions generation during earthquakes preparation

The well-known avalanche-like unstable model of fault formation describes the origination of different size cracks and finally, the main fault formation process in the earthquake focus (34).

Observations proved that when a rock is strained, electromagnetic emissions in a wide frequency spectrum diapason from MHz to kHz are produced by opening cracks (10,12,13,14,15,16,17).

Studies have shown that: 1) EM emissions appears about several weeks before the large earthquake; 2) The spectrum of electromagnetic radiation is characterized by the following sequence: MHz, kHz; 3) These radiations are accompanied by ULF radiation; 4) About 1-2 days before the earthquake, electromagnetic emissions become very weak or disappear at all (so-called "silence" appears). 5) The "silence" of EM radiation is followed by an earthquake (17,41).

The reason for the presence of ULF radiation during the earthquake
preparation process is completely understandable: in special scientific works are shown that the perturbations of the magneto-telluric field are caused by local and regional factors (31). The main fault formation process takes place in the conditions of the already perturbed telluric field (15, 41, 44).

Above mentioned observed experiences led us to the following:1) The radiating body must be in the focus without fail; 2) The regularity of the frequency change from MHz to kHz should indicate the stages of formation of radiated body i.e. main fault.

Based on these opinions, it has been developed the model of the generation of electromagnetic emissions (25). The work offers an interpretation of a mechanism for the formation of a hypothetical ideal electromagnetic contour in the focal zone of an incoming earthquake. The model of the generation of EM emissions is based on physical analogous to distributed and conservative systems and focal zones. According to the model, the process of earthquake formative from the moment of the appearance of cracks, to the completion of the whole process, including a series of foreshocks and aftershocks, can be entirely explained by the suggested oscillatory systems.

This work (25) offers a following formula:

$$\omega = k\frac{c}{l} \quad (1)$$

ω - self-generated frequency, $l$ - length of fault, $c$ - velocity of light.

Formula (1) connects with each other the main frequency of the observed electromagnetic emissions and the linear dimension of the emitted body (the length of the fault in the focus).

## Earthquake Prediction Methodology

In our view, if the electromagnetic radiation source, observed on the earth's surface during the earthquake preparation, is indeed the process of fault formation, there must be a harmonious coincidence between the two models mentioned above (25, 34) and the analysis of coincidence should give us certain opportunity to evaluate the characteristic parameters of the incoming earthquake.

The avalanche-like unstable model of fault formation is divided into three stages: the first stage can go on for several months throughout the whole seismogenic area. At this stage, the chaotic formation of micro-cracks without any orientation takes place. This stage of the formation of micro-cracks is the reversible process. Cracks created at this stage are small (34).

The short lengths of micro-cracks and the reversible process, according to our model (25), should be expressed by the discontinuous spectrum of electromagnetic radiation in MHz diapason, which is proved by scientific works (11,15, 41).

The second stage of the geological model of fault formation is an irreversible process of already somewhat oriented microstructures. Based on our model (25), this stage should be expressed by MHz continuous spectrum. Although, the values of electromagnetic emissions frequency must gradually decrease (because the lengths of the cracks start to increase at the expense of aggregation of primary small cracks). According to the geological model, this process takes place a few days before the main shock depending on the geological particularities (15, 41).

The transition of the MHz emissions in kHz in the frequency spectrum of electromagnetic radiation, according to formula (1), corresponds to the very stage when the crack length already reaches about a kilometer (15, 25, 27, 28, 41, 45).

At the final, third stage of the fault formation avalanche-like unstable process, the relatively big size faults unite into one - the main fault.

This process should correspond to the gradual fall of frequencies in kHz, which according to the formula (1) refers to the increase of fault length in the focus.

An increase of crack length in focus refers to the increase of magnitudes of the expected earthquake (48, 49):

$$lg\ l = 0.6\ M_s - 2.5 \quad (2)$$
$$M_w = 4.38 + 1.49 * \log l \quad (3)$$

$l$ - numerical value of the fault length in *km*. It must be noted that (2, 3) formulas are just for M≥5 earthquakes.

Of course, an association of cracks into one fault which at the final stage of earthquake preparation proceeds intensely, will use a definite part of energy accumulated in the focus and therefore, will result in its decrease (24).

In such a situation before the earthquake, (which can last from several hours to even 2 days), a fault is already formed but tectonic stress is not yet sufficient to overcome the limitation of the solidity of the environment (of course, at the approach of the critical value of tectonic stress, the balanced state in the system will be deranged and the earthquake will occur).

The system, which is waiting for a further "portion" of tectonic stress, is in the so-called "stupor - waiting" condition, that is, the process of fault formation is not going on in it anymore, and respectively, electromagnetic emissions would not take place. This is proved by experiments too (17).

Experiments prove the existence of "electromagnetic emissions silence" some hours (up to 2 days) before the earthquake (17).

Up to interruption of electromagnetic emissions, by using the final value of the frequency according to (1), we can determine, with rather the high accuracy, the fault length in the incoming earthquake focus that is, a magnitude (25, 48, 49).

It is the method of determination of one characteristic parameter (magnitude).

In seismology earthquake magnitude is measured only after the earthquake. Electromagnetic radiation has a great advantage because it arises during the earthquake preparation and gives the possibility to determine magnitude in advance of earthquake occurrence (11, 15, 16, 25, 27, 41, 48, 49).

Fixing the interruption moment of electromagnetic emissions is urgent for the determination of earthquake occurrence time since at the final stage of earthquake preparation, a very short time is needed to fill in the critical reserve of tectonic stress necessary for earthquake realization (24).

It is the possibility to determine the second parameter (earthquake occurrence time).

As for the determination of the expected earthquake epicenter (the third characteristic parameter), when the avalanche-like unstable process of fault formation will already be expressed in the EM emissions data, it is possible to define the epicenter of an incoming earthquake by the Direction-finding method with certain accuracy.

According to the model of EM emissions generation (25), the epicenter area should be approximately limited by positive electric potential. This theoretical conclusion is proved in cases of earthquakes in nature too (9). Obviously, bad weather or any technogeneous process can change potential sign of the earth's local area, but it is not a problem to filter the appropriate field.

Of course, the different interesting articles are devoted to the problem of epicenter determination (3, 6, 23, 26). However, above we represent the possibility of defining the epicenter of an incoming earthquake according to our theoretical model (25).

It is worth underlying that no reliable criterion is in seismology yet, which can distinguish large foreshocks from the main shock. This issue can be resolved with rather a high accuracy on the basis of a theoretical model (25): if after any shock electromagnetic emissions still continues to exist and the frequency data still tend to decrease, it means that the process of fault formation in the earthquake focus is not completed yet and we have to wait for the main shock.

The present methodology explains a general, that is, "classical" picture of earthquake preparation and occurrence (foreshock – main shock – aftershock) on the basis of the theoretical and fault formation avalanche-like unstable models (25, 34).

In case of weak earthquakes, we have to wait for the electromagnetic emissions in high frequency diapason but high-frequency waves attenuate rapidly, and to observe them on the earth's surface is difficult.

Besides, we should underline the fact that because relatively weak earthquakes do not pose a great danger, modern electromagnetic emissions networks are not focused on fixing corresponding frequencies.

## Earthquake Prediction Methods

The above-presented methodology has been checked on retrospective data of INFREP (European Network of Electromagnetic Radiation) for the Crete earthquake with M= 5.6 (25/05 / 2016, 08:36:13 UTC). Studies have been conducted by the data for 73 days (29).

INFREP network fixes every minute amplitudes of 10 different frequencies of VLF/LF electromagnetic radiation in diapason 20 270 Hz - 270 000 Hz.

According to our view, if any frequency channel actually reflects the earthquake preparation, the relevant geological process (34) should be reflected in the frequency data (formula 1).

For this reason, we calculated the lengths of every minute cracks corresponding to all frequency channels towards the lengths relevant to the channel's baseline frequencies (in the percentage) (25). It found out two active channels (37 500 Hz) and (45 900 Hz) as the average daily value of the cracks lengths were the maximum for them. This means that from discussed 10 channels, only two, with above mentioned frequencies, described the earthquake preparation process. In this case, according to the above given formulas, the magnitude of the incoming earthquake should be between 5.5 and 5.7 (Crete earthquake magnitude is really estimated as M= 5.6).

Besides, we found out that date diurnal periodic variations are clearly expressed on all channels, except the (37 500 Hz) channel. We have such variations in the F channel recordings too, but till some period, up to 02.05.2016, after which the anomalous process starts, indicating that the avalanche–unstable process of fault formation already began (Fig. 1).

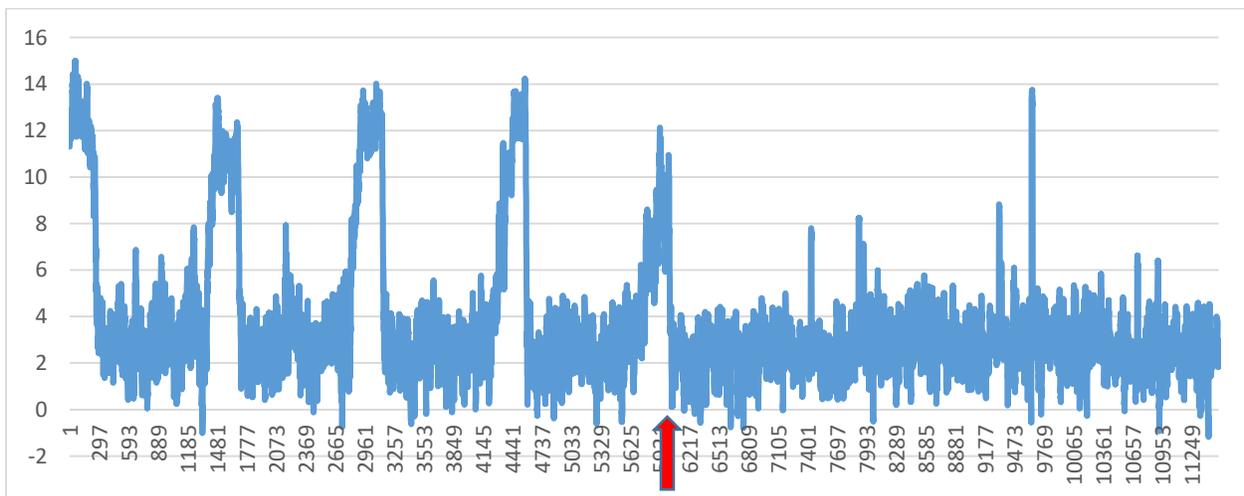

Fig.1. The initial moment of the avalanche-like unstable process of fault formation

Since, in case of discussed earthquake, only (37 500 Hz) frequency channel meets both conditions: for this channel the average daily value of the lengths of the cracks is maximal and an avalanche process of fault formation appears only on it, obviously, to predict the earthquake, we must rely only on the data of this channel.

During analyses, has been found, that 19 days before the earthquake, the fault formation avalanche process appeared in (37 500 Hz) frequency channel. By the formulas (1, 2, 3), the expected length of the main fault is about 8 000 meters, and the magnitude approximately is equal to 5.6.

In order to analysis of the full process of earthquake preparation, we elaborated the daily averaged frequencies by using the average square deviation method, and we calculated $\overline{x} \pm \sigma$ significances (fig.2).

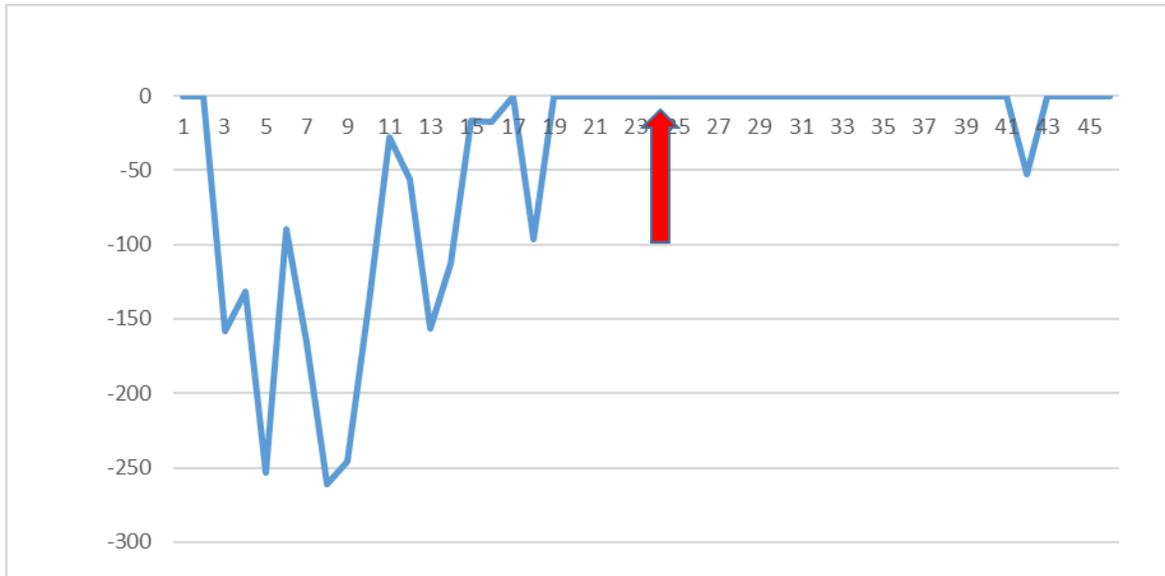

Fig. 2. The avalanche process of fault formation and EM emissions "silence" period before the considered earthquake.

Thus, the results of studies on real records of the electromagnetic field accurately described a major fault forming process in the focus of an incoming earthquake.

In results of the research the following conclusions are made:

1) By the active channel frequency, it is possible to determine the length of the "cracked strip" on which the process of cracks origination is going on actively and ultimately the main fault forms;

2) By the length of the "cracked strip", it is possible to determine the magnitude of an incoming earthquake with certain accuracy about several dozen days before the earthquake;

3) After the active frequency channel detection, it is already possible to determine the future earthquake epicenter with certain accuracy;

4) In order to the short-term prediction of a large earthquake, it is recommended to begin careful monitoring of the frequency data from the beginning of the fault formation avalanche-unstable process, keep an eye on the process dynamics, and fix the starting moment of "silence" of the electromagnetic radiation.

In the case of monitoring of electromagnetic radiation, it is possible to make a prediction of an incoming earthquake about from several dozen up two days before earthquake occurrence;

5) By the proposed method, it is easy to separate the foreshock from the mainshock.

Thus, VLF/LF EM emissions turned out to be the unique precursor, which gives promising possibility of simultaneous determination of large, inland incoming earthquake epicenter, time of occurrence, and magnitude. It is worth underlining that the revealed precursor is the first and only among other precursors, which describes the fault formation process in the incoming earthquake focus and numerically calculates fault length (magnitude) at any moment of monitoring.

## An earthquake precursor mobile network and accumulated tectonic stress reduction possibility

The existent networks employed to detect electromagnetic radiation consist of stationary transmitters and receivers. However, there are reported cases when networks are unable to detect any electromagnetic radiation before large earthquakes.

To avoid this blemish, we suggest the modernization of the existing networks by a network of mobile receivers without transmitters, based on the opinion that the perfect recording of radiated electromagnetic emissions depends on the orientation of the main fault towards the receiver (25, 29). First of all, it is necessary to conduct a geodetic survey for tectonic anomalies detection in the region, and then, the spatial distribution and configuration of the electromagnetic mobile receivers must be chosen (29).

Concerning the earthquake problem, we consider another important issue - the possibility of tectonic stress artificially reducing: at the end of the last century A.V.Nikolaev published article (40), where he notes that artificially initiating the growth of weak and moderate seismicity, accelerating the preparation process of a large earthquake, and as a consequence, reducing its magnitude should be considered possible.

Artificial discharge of tectonic energy can be done in two different ways: The first is the initiation of discharge of large areas by regular "processing" with powerful electrical and seismic sources, underground nuclear or chemical explosions. The second method is a purposeful impact on the focus of the incoming large earthquake. Powerful pulsed electrical sources can be replaced by sources of relatively low power, but long-acting. For purposeful impact it is necessary to find a place where nature has already prepared a catastrophic earthquake, i.e. to solve the problem of prognosis an earthquake (40).

We think, that today, based on our research (25,29) and by using special, reasonably organized impacts on the earthquake focus, it is possible to carry out a controlled discharge of tectonic energy and reduce seismic hazard (19, 30, 33, 37, 38, 39, 46, 47).
.